\documentclass[reprint,superscriptaddress,showpacs,nofootinbib,aps,prl]{revtex4-1}
\usepackage{amssymb}
\usepackage{amsmath}
\usepackage{graphicx}
\usepackage{dcolumn}
\usepackage{bm}
\usepackage{natbib}
\usepackage[utf8]{inputenc}
\usepackage[T1]{fontenc}

\begin{document}

\title{Thermodynamic model of\\
the macroscopically ordered exciton state}
\author{S. V. Andreev}
\email[Electronic adress: ]{Sergey.Andreev@univ-montp2.fr}
\affiliation{Laboratoire Charles Coulomb, Unite Mixte de Recherche 5221
CNRS/UM2, Universite Montpellier 2, Place Eugène Bataillon, 34095
Montpellier Cedex, France}

\begin{abstract}
We explain the experimentally observed instability of cold exciton gases and
formation of a macroscopically ordered exciton state (MOES) in terms of a
thermodynamic model accounting for the phase fluctuations of the condensate. 
We show that the temperature dependence of the exciton energy exhibits fundamental
scaling behavior with the signature of the second order phase transition.   
\end{abstract}

\pacs{Valid PACS appear here}
\maketitle

\preprint{APS/123-QED}

Long-range order and order parameter build up
are key features of Bose-Einstein condensation (BEC) in gases \cite%
{Pitaevskii}. All these features have been experimentally observed in cold
gases of indirect excitons in coupled semiconductor quantum wells (CQW) \cite%
{Butov2007}. Indirect excitons are formed by electrons and holes confined in
separate layers of CQW structure. Being bosonic quasi-particles, excitons
have quantum degeneracy temperatures several order of magnitude higher than
atoms \cite{Keldysh}, thus they are very attractive for studies of BEC. 

\begin{figure}[tbp]
\includegraphics[width=1\columnwidth]{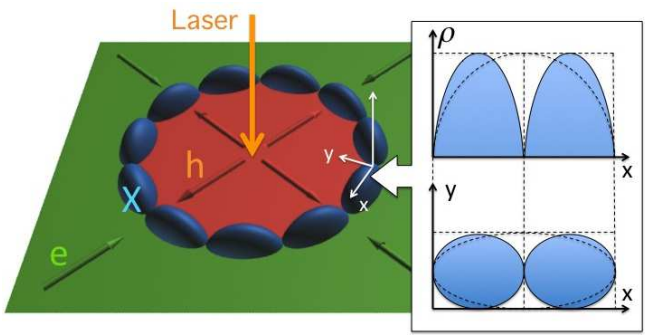}
\caption{Schematic view of the exciton density profile. Excitons (blue) are
created on the boundary between electron- and hole-ritch regions (green and
red). In the Thomas-Fermi limit the condensate density profile takes the
form of the inverted paraboloid having the height $\protect\rho _{max}=%
\protect\mu /V_{0}$ fixed by the chemical potential. The adiabatic topological
tranformation of a bead, shown in the inset, conserves the number of
particles and the energy of the system (see the main text).}
\label{fig1}
\end{figure}

Recently, the emergence of extended spontaneous coherence was observed at
low temperatures in a gas of indirect excitons cooled down to 100 mK \cite%
{High2012, nanoletters}. High \textit{et. al.} performed
shift-interferometry measurements of the off-diagonal one body density in
the external ring of the exciton photoluminescence pattern. The ring is
formed at the boundary between electron-rich and hole-rich regions \cite%
{formation}. Since this boundary is far from the laser excitation spot, the
ring is a source of cold excitons \cite{Butov2002, Rapaport}. The
experimentally observed photoluminescence pattern changes drastically at the
temperatures below 2K. The ring fragments into regularly spaced beads of
high PL intensity, having macroscopic sizes (Fig. \ref{fig1}).
Off-diagonal one-body density appears to be extended well beyond the thermal
de Broglie wavelength in the vicinity of one bead \cite{High2012}.

Levitov and co-workers \cite{Levitov} explained the transition of the
exciton system into this new macroscopically ordered state (MOES) in terms of a classical transport theory.
 An alternative explanation based on the existence of attractive
van der Vaals interactions between the excitons which might lead to
formation of the islands of electron-hole liquid was proposed simultaneously
in \cite{Sugakov}. Soon after Paraskevov \textit{et. al.} \cite{Paraskevov}
pointed out that both theories neglect the repulsive dipole-dipole exciton
interaction which is expected to play an important role in the process of
the bead formation \cite{Repulsive}. Levitov's system of coupled non-linear diffusion equations supplemented with the drift term due to Coulomb interactions was subsequently studied numerically in \cite{Wilkes}. 

The long-range order build up was explicitly taken into account in \cite{Paraskevov}. They showed that the spatially
nonuniform distribution of the condensate density can be obtained as a
standing wave type solution of the quasi-one dimensional Gross-Pitaevskii
equation. However, the recent studies \cite{High2012} have shown that there
is no coherence between different beads. The MOES is a fragmented condensate.
Similarly to atomic gases, two-body interactions
are expected to play significant role in the exciton condensates yielding to a ritch phenomenology \cite{Rontani, Shelykh, Rubo}.

In this context, there is an apparent need of theoretical description of a
system of multiple condensates (or a fragmented condensate) periodically
arranged on a ring. Here we find the critical conditions for condensate
fragmentation and describe the ground state of the system by means of a
thermodynamical consideration taking into account repulsive interactions between the excitons.
 We assume that the ring-like cloud of classical excitons
undergoes both condensation and fragmentation into beads with lowering of
temperature. The fragmentation, being purely quantum phenomena, can be regarded as the manifestation of the fundamental uncertainty principle for the phase and the particles number considered as canonically conjugated variabes. Our model shows that
formation of a fragmented condensate is driven by spatial fluctuations of the phase of the order parameter and the localization is achieved due to the increase of the entropy of the system. The model reproduces the experimental
dependencies of the number of beads on the pumping power and of the energy
of excitons on temperature.

At the densities achieved in the experiments on MOES \cite{Butov2002}, excitons can be viewed as weakly interacting bosons \cite{Rontani, Butov2007} and treated in the mean field approximation regardless the underlying band structure \cite{Snoke}. 
The macroscopic charge separation can produce an in-plane confinement for the excitons in the radial direction [SI], while the repulsive electrostatic interaction between the neighboring beads provides the azimuthal \textit{autolocalizing} potential.  Thus each bead can be considered as a two-dimensional condensate in a trap. The latter we will assume to be of a harmonic type. The relevant energy scale of the problem will be provided by the critical temperature $T_{c}$ of the BEC in a trap which is determined by the total number of excitons in the ring $N$ \cite{Pitaevskii} fixed by the gate voltage and the laser excitation power \cite{formation}. Note that the healing length characterizing the variation of the exciton density at the edges of the condensate \cite{Pitaevskii} is much smaller than the size of a bead [SI] and all quantities can be calculated in the Thomas-Fermi limit, neglecting the kinetic energy term in the Gross-Pitaevskii equation. In this case the density profile of each domain takes the form of the inverted paraboloid and the fragmentation of the condensate may be
achieved by an adiabatic transition \cite{footnote} conserving its potential energy $E$ and chemical potential $\mu$, which is set by the external reservoir. This essentially topological result can be understood from the schematic illustration in the inset of Fig. \ref{fig1}. The height of each paraboloid is fixed by the chemical potential according to the relation $\rho _{max}=\mu /V_{0}$, where $V_{0}$ is the interaction constant \cite{Baym}. We assume the transverse diameter of the base ellipse $w$ (the ring width) to be a constant in the first approximation. Now, if one devides the conjugated diameter of the initial paraboloid (dashed line) into $n$ parts and replaces it by $n$ similar paraboloids, then the integrals $\int d^{2}r\rho$ and $\int d^{2}r\rho ^{2}$ are conserved. The former (volume of the figure) is simply the total number of excitons $N$ and the latter, according to the 2D virial relation [SI], is related to the energy of a condensate by $E=V_{0}\int d^{2}r\rho ^{2}$.

Though the beads can in principle have different sizes (and, consequently, contain different number of excitons), the ground state of the system is expected to have a symmetric shape (Fig. \ref{fig1}), since this shape minimizes the kinetic energy in the small overlapping region between the neighboring condensates \cite{Pitaevskii, Boundary}. Thus, the fragmented condensate will be macroscopycally ordered in the thermodynamic equilibrium. Below we argue that the fluctuations of such density distribution are negligible and show that the transition of the condensate into this "number squeezed" configuration is driven by spontaneous breaking-up of the phase of the condensate, that increases the entropy of the system.

The transition from a coherent to an incoherent regime associated with the increase of the spatial fluctuations of the global phase of the condensate can be conveniently studied by means of quantized Josephson Hamiltonian, which in the $\Phi$-representation takes the form \cite{Fluctuations} 
\begin{equation}
\hat{H}_{J}=-\frac{1}{4}E_{C}\displaystyle\sum_{i=1}^{n}\frac{\partial^{2}}{\partial\Phi_{i}^{2}},
\label{Josephson}
\end{equation}
where $\Phi_{i}$ is the phase of the i-th fragment, $E_{C}=2\partial{\mu}/\partial{N_{0}}$ is an interaction parameter calculated at $N_{0}=N/n$ and $n$ is the number of fragments. The reduced form \eqref{Josephson} corresponds to the limit of no coherence between the beads. The eigenstates of the Hamiltonian \eqref{Josephson} are plain waves
\begin{equation*}
\Psi_{\lbrace k_{i}\rbrace}\sim\exp\Bigl(i\displaystyle\sum_{i=1}^{n}k_{i}\Phi_{i}\Bigr)
\end{equation*}
for set of integer values ${\lbrace k_{i}\rbrace}$, so that the ground state function is a constant, revealing that the phases of the beads are distributed in a random way. According to the uncertainty relation arising from the quantization of the Josephson equations \cite{Fluctuations, Carruthers, Leggett} the deviations $k_{i}$ of the number of particles in coherent state in each site from their equilibrium values $N_{0}=N/n$ are instead vanishingly small - the fragments have well-defined number of excitons (Fock state).

The set of the eigenvalues of the Hamiltonian \eqref{Josephson} is given by
\begin{equation}
E_{\lbrace k_{i}\rbrace}=-\frac{E_{C}}{4}\displaystyle\sum_{i=1}^{n}k_{i}^{2}
\label{Spectrum}
\end{equation}
and determines the spectrum of elementary excitations associated with the phase dynamics of the condensate. 
One can see that the break-up of the global phase leads to the appearence of new degrees of freedom. The partition function of the system can be factorized and it takes the form
\begin{equation}
Z_{\Phi}\equiv\displaystyle\sum_{\lbrace k_{i}\rbrace}e^{-\beta E_{\lbrace k_{i}\rbrace}}=\Bigl(\sum_{k}e^{-\beta E_{k}}\Bigr)^{n},
\label{Z}
\end{equation}
where $E_{k}=E_{C}k^{2}/4$. Using Eq. \eqref{Z} one can straightforwardly evaluate [SI] the entropy
\begin{equation}
S_{\Phi}=\frac{nk_{B}}{2}\Bigl[1+\ln\Bigl(\frac{4\pi}{\eta}\frac{T}{T_{c}}\frac{N}{n}\Bigr)\Bigr]
\label{entropy}
\end{equation}
and the energy
\begin{equation}
E_{\Phi}=\frac{nk_{B}T}{2}
\label{energy}
\end{equation}
of the system due to the excitations \eqref{Spectrum}, where we have substituted 
$E_{C}=\mu/N_{0}$
holding in the 2D Thomas-Fermi limit [SI]. The parameter $\eta\equiv\mu/k_{B}T_{c}$ characterizes the strength of interactions as discussed below (see Eq. \eqref{eta}). Substitituting the expressions \eqref{entropy} and \eqref{energy} to the canonical potential $F_{\Phi}=E_{\Phi}-TS_{\Phi}$ one can see that the latter decreases as function of the number of beads $n$ while $n\ll N$. Therefore, an unfragmented state is thermodynamically unstable and the ring will irreversibly break up increasing its entropy $S_{\Phi}$. To find the steady state one needs to go beyond the Thomas-Fermi limit. The upper limit for the number of beads $n$ would be imposed by the increase of the kinetic energy due to localization which we have neglected so far. These "quantum pressure" corrections arise from the small region near the boundary between the adjacent beads, which is not correctly accounted for in the Thomas-Fermi limit. Its proper description requires the explicit inclusion of the quantum effects in the Gross-Pitaevskii equation \cite{Boundary}. Here we phenomenologically introduce the relevant energy correction for each bead as $\sigma$. It can be considered as an energy of a boundary separating two beads. It can be shown [SI], that $\sigma=xk_{B}T_{c}/\eta$, where $x$ is a numerical coefficient defined by the topology of the condensate in the boundary region between the neighboring fragments. Accounting for this correction, the free energy of the ring then takes the form
\begin{equation}
F=n\frac{x}{\eta}k_{B}T_{c}-\frac{nk_{B}T}{2}\ln\Bigl(\frac{4\pi}{\eta}\frac{T}{T_{c}}\frac{N}{n}\Bigr),
\label{F}
\end{equation}          
and, considered as a function of $n$, has a minimum at the point
\begin{equation}
n_{0}=\frac{4\pi}{\eta e}tN\exp\Bigl(-\frac{2x}{\eta}t^{-1}\Bigr),
\label{n0}
\end{equation}
where $t\equiv T/T_{c}$ is the reduced temperature. Eq. \eqref{n0} gives the number of the beads in the steady state of the fractioned condensate on a ring in a simple form of the
Arrhenius activation law \cite{Arrhenius} and demonstrates the crucial role of the two-body interactions in the process of formation of MOES. Only if the interactions are significant ($\eta\gtrsim 0.2$) [SI] can the quantum fluctuations of the phase of the order parameter be sufficiently large to drive the system into the number squeezed state. The estimate for $\eta$ given below shows that this condition is indeed well satisfied in our case. 

The result \eqref{n0} can be
directly compared with the experimental data on the dependence of the number
of beads on the pumping intensity $P$. Following Snoke \textit{et. al.} \cite{SnokeNature} we will assume the total mean number of excitons in the steady state depending linearly on the laser excitation power at the fixed gate voltage. We argue that if, in addition, the \textit{ring radius} also depends linearly
 on $P$, then the critical temperature $T_{c}$ would be $P$-independent [SI]. Linear dependence of the ring radius on the pumping intensity has been indeed observed
experimentaly (see the inset in Fig. \ref{fig3}) and explained theoretically
within the kinetic model of the ring formation \cite{Haque}. The result of
the fitting using the formula \eqref{n0} is presented in Fig. \ref{fig3}. We find $x\sim 1$ that is consistent with an estimate for the quantum pressure done in \cite{Boundary}.

\begin{figure}[tbp]
\includegraphics[width=1\columnwidth]{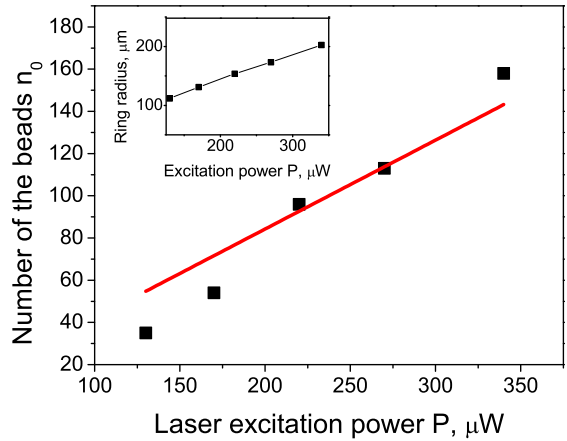}
\caption{The number of the beads in a steady state versus laser excitation
power. Red line is a fit to the experimental data (squares) using Eq. 
\eqref{n0}, where we substitute $N_{con}=N(1-t^{2})$, $N=\protect\beta P$ for the total number of excitons as a
function of the excitation power $P$. We take $\protect\beta%
=4150$ $\protect\mu W^{-1}$ which corresponds to the average exciton density in a bead $\bar\rho=10^{10}$ $cm^{-2}$ and gives a theoretical estimate [SI] for $T_{c}$ close to 4.5 K observed experimentally (see Fig. \ref{fig3}). $T_{c}$ is assumed to be P-independent providing that the ring radius increases linearly with $P$, as indeed observed experimentally (see the inset). The bath temperature $T=2$ K. From the fitting we deduce $x=3.2$ (see the main text).}
\label{fig2}
\end{figure}

So far we assumed sufficiently low bath temperature, so that one could neglect thermal depletion of the condensate and temperature dependence of the chemical potential.
Now we are going to extend our model to higher temperatures in order to
explain the nonmonotonic temperature dependence of the exciton energy
observed in \cite{Repulsive}. The crucial parameter of the problem will be
the ratio $\eta$ between the value of the chemical potential calculated
using the Thomas-Fermi approximation at $T=0$ and the critical temperature $%
T_{c}$ for noninteracting particles \cite{Giorgini}. It can be expressed as
[SI] 
\begin{equation}
\eta =\sqrt{\frac{\pi }{6}\frac{mV_{0}}{\hbar ^{2}}}.
\label{eta}
\end{equation}
The value of the interaction constant $V_{0}$ can be estimated using the
well-known plate capacitor formula corrected by a factor dependent on the
distance $d$ between the centers of the coupled quanum wells \cite%
{zimmermann}. For $d=12$ nm we obtain $\eta =1.6$.

The total energy of the system $E$ at $T\leqslant T_{c}$ is a sum of the
condensate energy $E_{con}$ and the energy of uncondensed excitons (thermal
component) $E_{th}$. The energy of the condensate can be calculated by
integrating the thermodynamic relation 
\begin{equation*}
\frac{\partial {E_{con}}}{\partial {N_{con}}}=\mu -T\frac{\partial {\mu }}{%
\partial {T}}.
\end{equation*}%
The temperature dependence of the chemical potential $\mu $ in the first
approximation can be obtained by substituting the estimate for the number of
excitons in the condensate $N_{con}=N(1-t^{2})$ obtained in the
non-interacting limit into the 2D Thomas-Fermi expression [SI] 
\begin{equation}
\mu (t)=k_{B}T_{c}\eta \Bigl(\frac{N_{con}}{N}\Bigr)^{1/2},  \label{muT}
\end{equation}%
where $t=T/T_{c}$ is the reduced temperature. One finds 
\begin{equation}
\frac{E_{con}}{Nk_{B}T_{c}}=2\eta (1-t^{2})^{1/2}.  \label{Econ}
\end{equation}%
In what concerns the uncondensed excitons, at $T\leqslant T_{c}$ they can be
treated as free particles propagating in the effective mean field potential $%
V_{eff}(x,y)-\mu (t)=\lvert V_{ext}(x,y)-\mu (t)\rvert $, which coincides
with the trapping potential $V_{ext}$ outside the condensate and is
drastically changed inside where it becomes repulsive \cite{thermodynamics}.
One can calculate the energy of the thermal component using the Bose
functions \cite{Pitaevskii}.  Using the expression \eqref{muT} for $\mu (t)$
one can find 
\begin{equation}
\frac{E_{th}}{Nk_{B}T_{c}}=\frac{2}{\zeta (2)}t^{3}g_{3}[\exp (-\eta
t^{-1}(1-t^{2})^{1/2})],  \label{Eth}
\end{equation}%
where $g_{3}(z)$ is a Bose function, in which the chemical potential $\mu $
is replaced by $(V_{eff}(x,y)-\mu (t)$). Summing the results \eqref{Econ}
and \eqref{Eth} we find the total energy of the system below $T_{c}$: 
\begin{equation}
\frac{E}{Nk_{B}T_{c}}=\frac{2}{\zeta (2)}t^{3}g_{3}[\exp (-\eta
t^{-1}(1-t^{2})^{1/2})]+2\eta (1-t^{2})^{1/2}.  \label{Ebelow}
\end{equation}

Above $T_{c}$, the system is very dilute and can be considered as an ideal
gas placed into the confining potential $V_{ext}$. Following the general
rules of statistical mechanics we derive the total energy of the ring for $%
T>T_{c}$ in the form 
\begin{equation}
\frac{E}{Nk_{B}T_{c}}=\frac{2}{\zeta (2)}t^{3}g_{3}(z),  \label{Eabove}
\end{equation}%
where $z$ is a root of the equation $g_{2}(z)=\zeta (2)t^{-2}$.
\begin{figure}[tbp]
\includegraphics[width=0.6\columnwidth]{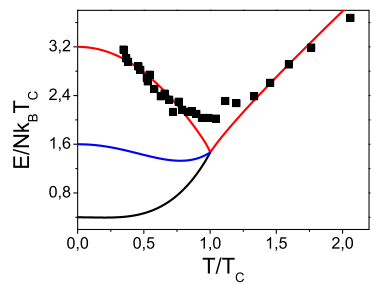}
\caption{The exciton energy in units $k_{B}T_{c}$ versus the reduced
temperature $t=T/T_{c}$. Solid lines in the region $t<1$ (below $T_{c}$) are
the result of calculation using the formula \eqref{Ebelow} for $\protect\eta%
=0.2$ (black), 0.8 (blue) and 1.6 (red). The result of the calculation above 
$T_{c}$ (the region $t\geqslant 1$) using the ideal gas model \eqref{Eabove}
is independent on $\protect\eta$ and is presented by red line. Experimental
data are taken from \protect\cite{Repulsive}. The critical temperature $%
T_{c}=4.5$ K. The specific heat at constant volume $C_{V}=\partial{E}%
/\partial{T}$ exhibits the discontinuity at the critical point.}
\label{fig3}
\end{figure}

Eqs. \eqref{Ebelow} and \eqref{Eabove} are expressed in terms of only two
parameters ($\eta $ and $t$), which reflects the fundamental \textit{scaling
behavior} exhibited by the system in the limit of large $N$. The scaling
behaviour of condensates is also well-known for the 3D case \cite%
{thermodynamics}. The temperature dependence of the exciton energy in units
of $k_{B}T_{c}$ for different values of $\eta$ is plotted in Fig. \ref{fig3}. It shows the non-monotonic behaviour with a minimum corresponding to the
critical temperature. The best agreement with the experimental data is
achieved at $\eta =1.6$, which fits excellently to the value of $\eta$
calculated above from the microscopic model \cite{zimmermann}. Note, that
the specific heat at constant volume $C_{V}=\partial {E}/\partial {T}$
exhibits a discontinuity at $T=T_{c}$. This the signature of the second
order phase transition.

To conclude, we have presented a thermodynamical model of formation of a
macroscopically ordered exciton state. It shows that the transition of a ring-like exciton condensate into the number squeezed fragmented state is driven by spatial fluctuations of the phase of the condensate.
The steady state of the system is determined by the balance between the kinetic energy and the entropy. Minimizing the free energy yields the number of the beads on the ring which depends on the reduced temperature following the Arrhenius activation law. The method allows tracing the exciton energy as a function of temperature as well. Both dependencies exhibit the characteristic scaling behaviour. The excellent
agreement of the calculated exciton energies with the experimental data \cite%
{Repulsive} confirms \textit{aposteriori} the presence of the second order phase transition in the exciton system.

We thank L. Butov and J. Leonard for granting us access to the unpublished experimental
results. The author is also indebtful to A. Kavokin, K. Kavokin, Yu. Rubo and M. Dyakonov for valuable discussions. This work has been supported by a EU ITN project "CLERMONT 4".


\begin{thebibliography}{99}
\bibitem{footnote} The entropy of the thermal component of the exciton gas will not be affected by the change of the effective mean field potential as long as $\tau\gg\omega_{ho}^{-1}$, where $\tau$ is the fragmentation time scale and $\omega_{ho}$ is the oscillator frequency of the trap \cite{Landau}.
\end{thebibliography}
\end{document}